\documentclass[conference]{IEEEtran}
\IEEEoverridecommandlockouts
	
\usepackage{cite}
\usepackage{amsmath,amssymb,amsfonts}
\usepackage{algorithmic}
\usepackage{graphicx}
\usepackage{textcomp}
\usepackage{xcolor}
\usepackage{breqn}
\def\BibTeX{{\rm B\kern-.05em{\sc i\kern-.025em b}\kern-.08em
    T\kern-.1667em\lower.7ex\hbox{E}\kern-.125emX}}
\begin{document}    
\title{Pan-Tompkins++: A Robust Approach to Detect R-peaks in ECG Signals}
\author{\IEEEauthorblockN{1\textsuperscript{st} Md Niaz Imtiaz}
\IEEEauthorblockA{\textit{Electrical, Computer, and Biomedical Engineering} \\
\textit{Toronto Metropolitan University}\\
Toronto, Canada \\
niaz.imtiaz@ryerson.ca}
\and
\IEEEauthorblockN{2\textsuperscript{nd} Naimul Khan}
\IEEEauthorblockA{\textit{Electrical, Computer, and Biomedical Engineering} \\
\textit{Toronto Metropolitan University}\\
Toronto, Canada \\
n77khan@ryerson.ca}}
\maketitle
\begin{abstract}
R-peak detection is crucial in electrocardiogram (ECG) signal processing as it is the basis of heart rate variability analysis. The Pan-Tompkins algorithm is the most widely used QRS complex detector for the monitoring of many cardiac diseases including arrhythmia detection. However, the performance of the Pan-Tompkins algorithm in detecting the QRS complexes degrades in low-quality and noisy signals. This article introduces Pan-Tompkins++, an improved Pan-Tompkins algorithm. A bandpass filter with a passband of 5–18 Hz followed by an N-point moving average filter has been applied to remove the noise without discarding the significant signal components. Pan-Tompkins++ uses three thresholds to distinguish between R-peaks and noise peaks. Rather than using a generalized equation, different rules are applied to adjust the thresholds based on the pattern of the signal for the accurate detection of R-peaks under significant changes in signal pattern. The proposed algorithm reduces the False Positive and False Negative detections, and hence improves the robustness and performance of Pan-Tompkins algorithm. Pan-Tompkins++ has been tested on four open source datasets. The experimental results show noticeable improvement for both R-peak detection and execution time. We achieve 2.8\% and 1.8\% reduction in FP and FN, respectively, and 2.2\% increase in F-score on average across four datasets, with 33\% reduction in execution time. We show specific examples to demonstrate that in situations where the Pan-Tompkins algorithm fails to identify R-peaks, the proposed algorithm is found to be effective. The results have also been contrasted with other well-known R-peak detection algorithms. Code available at: \url{https://github.com/Niaz-Imtiaz/Pan-Tompkins-Plus-Plus}
\end{abstract}


\section{Introduction}
\label{intro}
Cardiovascular diseases are the major cause of mortality worldwide, with nearly 17.8 million deaths per year. The most commonly utilized cardiovascular disease diagnostic technique is the electrocardiogram (ECG).  Worldwide, approximately 3 million ECGs are generated every day.  As wearable technology advances, the number of ECGs generated for examination only increases. Automated diagnosis techniques are necessary to analyze ECGs produced by wearable technology and to lessen the workload for doctors.
An electrocardiogram (ECG) is a visual representation of the electrical signals produced during a cardiac cycle. It gives details about the rhythm and morphology of heart rate. Cardiologists utilize ECG to monitor heart conditions and to detect cardiac illnesses such as arrhythmias, hyperkalemia, and myocardial infarction. A large amount of medical information can be obtained by measuring the amplitudes and durations of the ECG signals. In order to compute heart rate and detect arrhythmias, R-peak identification in ECG signal is crucial.

The ECG signal is made up of various waves, including P-waves, QRS complexes, and T-waves. A QRS complex is a brief-duration pulse that contains an R-peak with a high amplitude. The majority of automatic computer-based solutions concentrate on identifying QRS complexes \cite{rizwan2022machine}. Clinically, the accurate identification of the QRS complex is essential. Any deviation from the typical ECG sinus rhythm caused by the atria and ventricles depolarizing or repolarizing prematurely or slowly results in a change in the morphology of the QRS, which is further linked to one or more cardiac disorders \cite{gupta2021bp}.

 The most common technique for QRS detection in commercial devices is the one proposed by Pan and Tompkins \cite{pan1985real}. With accurate clinical ECG signal data, this approach offers good detection performance. However, low-quality and noisy signals, as well as ambulatory ECG recordings, reduce the Pan-Tompkins algorithm's capacity to detect QRS complexes \cite{fariha2020analysis}. This study analyzes the limitations of the Pan-Tompkins algorithm. We analyze the False Positives and False Negatives given by the Pan-Tompkins, identify the causes behind them, and introduce an efficient algorithm, namely Pan-Tompkins++, that boosts the performance of the original algorithm. 
 
We applied a bandpass filter with a passband of 5–18 Hz to remove noise without discarding significant signal components. An N-point moving average filter has been applied to smooth the signal. To further deal with noise, three thresholds have been used. Instead of adjusting them with a generalized equation, they are adjusted by different rules based on signal pattern so that the correct detection of R-peaks is possible even with a change in the pattern. We tested our proposed algorithm on four open-source ECG datasets- MIT-BIH Arrhythmia Dataset, St.-Petersburg Institute of Cardiological Technics 12-lead Arrhythmia Dataset, European ST-T Dataset, and Computing in Cardiology Challenge 2014 Dataset \cite{goldberger2000physiobank}\cite{moody2001impact}\cite{taddei1992european}. We compare our method with Pan-Tompkins \cite{pan1985real}, Hamilton \cite{hamilton2002open}, Engzee modified by  Lourenco \cite{engelse1979single}\cite{lourencco2012real}, Wavelet Transform \cite{kalidas2017real}, and Two Moving Averages \cite{elgendi2010frequency}. 

\section{Related works}

Over the years, numerous approaches have been proposed to detect QRS complexes and R peaks. Hamilton \cite{hamilton2002open} used low-pass and high-pass filtering to remove noise, differentiation to obtain the QRS slope information, and moving average filtering to produce information about the width and slope of the QRS complex. A detection threshold is used to differentiate beat and noise. Engelse and Zeelenberg \cite{engelse1979single} suggested a single scan approach to detect QRS and extract features. In their approach, a digitally filtered ECG signal is fed via a differentiator and a low pass filter afterward. The processed signal's negative lobes are examined in order to identify the R peak. They employed fixed thresholds for identifying R peaks. If the experimented ECG signal involves amplitude changes, the fixed thresholds have a negative impact on the robustness of the algorithm. The algorithm was further modified by Lourenco et al. in 2012 \cite{lourencco2012real}. Instead of fixed thresholds, they employed adaptive thresholds.  Elgendi et al. \cite{elgendi2010frequency} noted that the accuracy of any QRS detection system is highly dependent on the frequency range of the ECG being analyzed. They introduced a sensitive QRS detection method and evaluated the effectiveness by employing various frequency bands. Bandpass filtering, creating possible blocks, and thresholding are the three key phases of their method. With the help of Stationary Wavelet Transforms (SWT), Kalidas and Tamil \cite{kalidas2017real} presented an online QRS detector technique for detecting beats from single-lead electrocardiogram (ECG) readings in real-time. The thresholds are initially set based on the first ten seconds of the ECG signal and then adjusted every three seconds for speedy adaptation to the variations in heart rate and signal quality. Silmane and Nait-Ali \cite{slimane2010qrs} introduced a QRS complex detection algorithm based on Empirical Mode Decomposition. Their approach is made up of multiple phases, including high-pass filtering, empirical mode decomposition, nonlinear transformation, and integration, and then a first-order low-pass filter is used to generate a distinct maximum for every QRS complex. 

Recently some machine learning-based approaches have been proposed for R-peak detection. Cai and Hu \cite{cai2020qrs} utilized two multi-dilated convolutional block-based deep learning models. One model, the CNN, is made up of convolutional blocks and squeeze-and-excitation networks (SENet). In the other model (CRNN), a convolutional and recurrent neural network are combined. Laitala et al. \cite{laitala2020robust} proposed a novel R-peak identification technique using the Long Short-Term Memory (LSTM) network. 

Although machine learning and Fourier or wavelet-based techniques can provide promising results, they are computationally expensive. For real-time heart monitoring applications using wearable devices, it is important to detect R-peaks in a computationally inexpensive way. Hence, morphological analysis-based techniques are most widely used.

\section{Proposed Method}
\label{model}
 Our proposed algorithm aims to improve the Pan-Tompkins algorithm. Therefore, we discuss the algorithm briefly first.
\subsection{Pan-Tompkins Algorithm}
\label{pan-tompkins}
Typically, the Pan-Tompkins algorithm is employed for detecting QRS complexes in real-time. To locate the R-peaks in QRS complexes, this approach makes use of the slope, amplitude, and width of an integrated window \cite{pan1985real}. The two phases of the algorithm are pre-processing and decision-making. Pre-processing involves noise elimination, signal smoothing, and enhancing width and QRS slope. The signal goes through a digital bandpass filter, which is made up of cascading high-pass and low-pass filters. The signal is then differentiated in order to examine the information about the QRS complex slope. This is done by a five-point derivative with the following transfer function:
\begin{align}
H(z)=(1/8T)({-z}^{-2}-2{z}^{-1}+2{z}^{1}+{z}^{2})
\end{align} 

In the derivative procedure, the low-frequency P- and T- waves are suppressed to gain the high-frequency signals existing in the steeper slopes of the QRS complex. The signal is then squared point by point. This makes every data point positive and nonlinearly amplifies the derivative's output to emphasize higher frequencies. This lessens the typical larger T-wave amplitudes that can lead to false detection. The Moving Window Integration (MWI) is applied to extract information from the waveform feature, with the R wave slope taken into account. It is calculated by the following equation:
\begin{dmath}
y(nT)=(1/N)[x(nT-(N-1)T)+x(nT-(N-2)T)+...+x(nT)]
\end{dmath} 
where, $N$ is the number of samples within the integration window's width.

The determination of the width of the moving window is important. If the window is too large, the QRS and T complexes will be merged in the integration waveform. If it is too small, some QRS complexes will generate several peaks in the integration waveform. A window width of 150ms has been used as it is found to be efficient.

The decision-making phase is carried out to determine whether or not the MWI result is a QRS complex. To guarantee that the proper peak was picked, two thresholds are applied. The first examination of the signal is conducted using the higher of the two thresholds ($Threshold_1$). If no QRS is found within a predetermined window of time, the lower threshold ($Threshold_2$) is applied, necessitating the deployment of a search-back approach to search backward in time for the QRS complex. The thresholds are adjusted to optimize the identification of the QRS complex. The thresholds are computed as per the following equations:
\begin{align}
\label{eqn:label3}
SPK= 0.125 \ PEAK +0.875 \ SPK
\end{align} 
\begin{align}
\label{eqn:label4}
NPK= 0.125 \ PEAK +0.875 \ NPK
\end{align} 
\begin{align}
Threshold_1 = NPK + 0.25 \ (SPK-NPK)
\end{align} 
\begin{align}
Threshold_2 =0.5 \ Threshold_1
\end{align} 

where,

$PEAK$ is the overall peak

$SPK$ is the running estimate of the signal peak

$NPK$ is the running estimate of the noise peak

\subsection{Pan-Tompkins++}
\label{proposedmodel}
The proposed algorithm aims to resolve some particular limitations of the Pan-Tompkins algorithm. The modifications have been made in both the pre-possessing and decision-making steps. We discuss each limitation and the proposed solution in the following sections. 
\subsubsection{Pre-processing}
\label{preprocessing}
In reality, ECG signals are frequently contaminated by many sorts of noise, including baseline wander, power-line interference, muscle artifacts, and electrode contact noise, making it difficult to conduct a morphological analysis \cite{joshi2013survey}. The frequency spectrum of the baseline wandering falls within the 0.05–1 Hz region \cite{satija2018review}. The power-line interference is a narrowband noise with a 50/60 Hz frequency center. Typically, the muscle artifact noise is roughly 10\% of the ECG amplitude with a bandwidth range of 20–1000 Hz. Previous studies have demonstrated that muscle artifacts can drastically affect the shapes of local waves as the noise largely overlaps with the ECG signal in the 0.01-100Hz range \cite{clifford2006ecg}.

\textbf{Limitation 1: Loss of significant signal components:}
\label{limitation1}
In the Pan-Tompkins algorithm, the low-pass and high-pass filters are cascaded to achieve around 5–15 Hz passband. However, some significant ECG signal components are lost.

\textbf{Proposed solution for limitation 1:}
\label{solution1}
We opt to utilize a bandpass filter with a passband of 5–18 Hz. This preserves the significant signal components within 15–18 Hz. 

\textbf{Limitation 2: Presence of random noise:}
\label{limitation2}
Presence of random noise degrades the performance of the Pan-Tompkins algorithm. 

\textbf{Proposed solution for limitation 2:}
\label{solution2}
An N-point moving average filter has been used to smooth the squared signal before it goes through the Moving Window Integration (MWI) step. A window with a 60ms width has been used. The smoothed signal is computed in this approach by convolution of the input signal with a filter kernel. We applied several commonly used window functions and noted the performance. The functions that have been tested are  Hamming, Kaiser, Blackman, flattop, Bohman, Gaussian, triangular, and rectangular windows. The flattop window performed the best of all of them. We select the flattop window in our algorithm for the smoothing operation. A window with a flat top minimizes scalloping loss, which is related to the main lobe's flatness in the frequency domain \cite{das2020implementation}. It is a window with partially negative values. The flattop window is defined as follows in the time domain:
\begin{dmath}
{\omega}(n)={a}_0 - {a}_1 {\cos}({\psi}) +{a}_2 {\cos}(2{\psi}) -{a}_3 {\cos}(3{\psi}) +{a}_4 {\cos}(4{\psi})
\end{dmath}
where,
$${\psi}=\frac{(2n{\pi})}{N} $$
$N$ = Number of samples within the width of the window function
$${a}_0=0.2155789; {a}_1=0.4166316; {a}_2=0.27726316;$$ $${a}_3=0.08357895; {a}_4=0.00694737 $$
\begin{figure}[htbp]
\begin{center}
    \includegraphics[height=6cm,width=8cm]{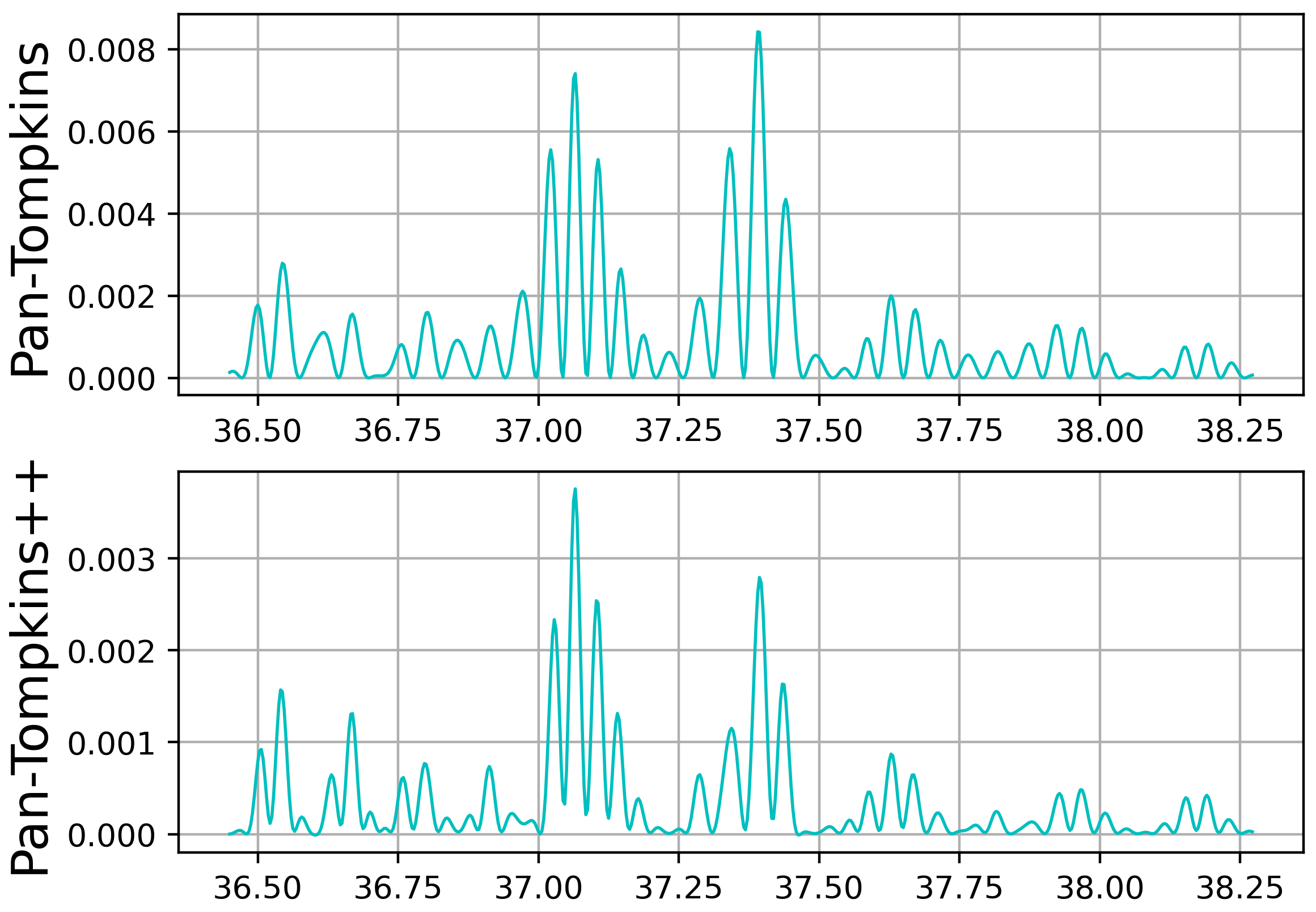}
    \caption{Presence of random noise and effectiveness of the moving average filter.}
\end{center}
\label{fig1}
\end{figure}
Figure 1 shows that the squared signal still contains some random noise. Unusual high magnitudes are frequently visible as a result (Figure 1: Pan-Tompkins). The random noise is eliminated during the smoothing operation (Figure 1: Pan-Tompkins++).  

\subsubsection{Decision Making}
\label{decisionmaking}
We use two sets of thresholds ($Threshold_1$ and $Threshold_2$) that are adjusted during every examination for an R-peak and one additional threshold ($Threshold_3$) that is only calculated while searching for a missing R-peak. We apply one set of thresholds on the filtered signal given by the bandpass filter and the other one on the signal after differentiation, smoothing, and moving window integration operations. 

We select all the peaks that are at least 231ms apart from each other. Hence, we allow a maximum heart rate of 260 bpm (beats per minute). Thereafter, we classify each peak as a noise peak or an R-peak. To qualify as an R-peak, the peak must exceed the higher threshold ($Threshold_1$) as the signal is first examined or the lower threshold ($Threshold_2$) if search-back is necessary to locate the QRS.

The thresholds are initialized by the following equations:
\begin{align}
Threshold_1 = MAXF/3 
\end{align}
\begin{align}
Threshold_2 = 0.5\ MEANF
\end{align}
\begin{align}
SPK= Threshold_1 
\end{align}
\begin{align}
NPK= Threshold_2 
\end{align}

where,

$MAXF$ is the maximum amplitude in the first 2s interval of the signal

$MEANF$ is the average amplitude in the first 2s interval of the signal

$SPK$ is the running estimate of the signal peak

$NPK$ is the running estimate of the noise peak

We define two rules ($Rule-1$ and $Rule-2$) for the adjustment of $SPK$ and $NPK$ to accurately identify R-peaks under
significant changes in the signal pattern. $Rule-1$ is exactly the same as the Pan-Tompkins algorithm (eqs. 3 and 4). $Rule-2$ is defined as follows:
\begin{align}
SPK= 0.75 \ PEAK +0.25 \ SPK
\end{align}
\begin{align}
NPK= 0.75 \ PEAK +0.25 \ NPK
\end{align}

The thresholds are adjusted after each examination for an R-peak. $Threshold_1$ is adjusted in the same way as Pan-Tompkins does (eq. 5), but $Threshold_2$ is adjusted as:
\begin{align}
Threshold_2 =0.4 \ Threshold_1
\end{align}
\textbf{Limitation 3: Skipping low amplitude R-peaks:}
\label{limitation3}
In the Pan-Tompkins algorithm, the lower threshold is calculated as half of the higher threshold. That is still a large value. Because of the large value of the lower threshold, we see many missing beats.

\textbf{Proposed solution for limitation 3:}
\label{solution3}
We adjust the lower threshold by multiplying the higher threshold by 0.4 (eq. 14). Figure 2 shows that, the Pan-Tompkins algorithm cannot detect the R-peaks (marked with a rectangular box) with lower amplitude. The modified adjustment of the lower threshold resolves this.

\begin{figure}[htbp]
\begin{center}
    \includegraphics[height=6cm,width=8cm]{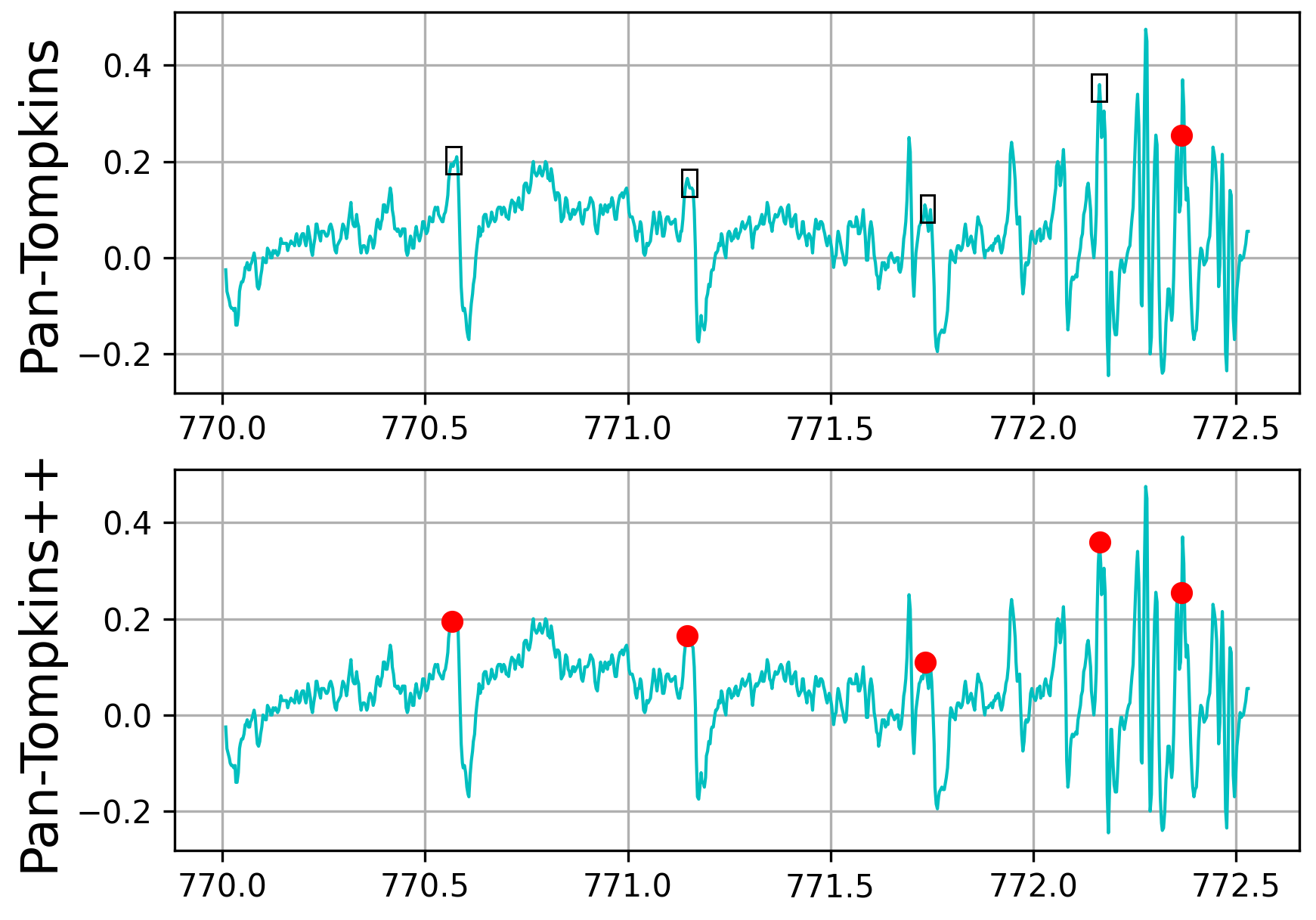}
    \caption{Missing low amplitude R-peaks by Pan-Tompkins (highlighted by black rectangles) and effect of proposed solution.}
\end{center}
\label{fig2}
\end{figure}

\textbf{Limitation 4: False beat detection after a real beat:}
\label{limitation4}
The Pan-Tompkins algorithm occasionally identifies R-peaks that are very close to one another. They are very often False Positives. 

\textbf{Proposed solution for limitation 4:}
\label{solution4}
RR-interval averages are maintained throughout the examination. This is the mean RR-interval of the eight most recent beats. When the current RR-interval is less than 360ms or it is less than half of the current RR-interval average, an assessment is done to determine whether the current QRS complex has been accurately diagnosed or whether it is actually a T-wave. This is done to avoid the False Positives. The RR-interval check is added as the Pan-Tompkins algorithm occasionally picks up false beats close to a real beat. The mean slope is calculated over the preceding 70ms window of both the current peak and the most recent detected R-peak. If the slope that occurs during the current waveform is less than 60 percent of that of the preceding QRS waveform, it is classified as a T-wave; otherwise, it is a QRS complex. If a QRS complex is found, the $SPK$ and $NPK$ are adjusted  using $Rule-1$ (eqs. 3 and 4). This avoids the false beats detected by the Pan-Tompkins algorithm. Figure 3 demonstrates how the Pan-Tompkins algorithm may identify false beats (marked with a black rectangle) that appear just after (within 360ms) the actual beats. The proposed algorithm is able to resolve that issue. 

\begin{figure}[htbp]
\begin{center}
    \includegraphics[height=6cm,width=8cm]{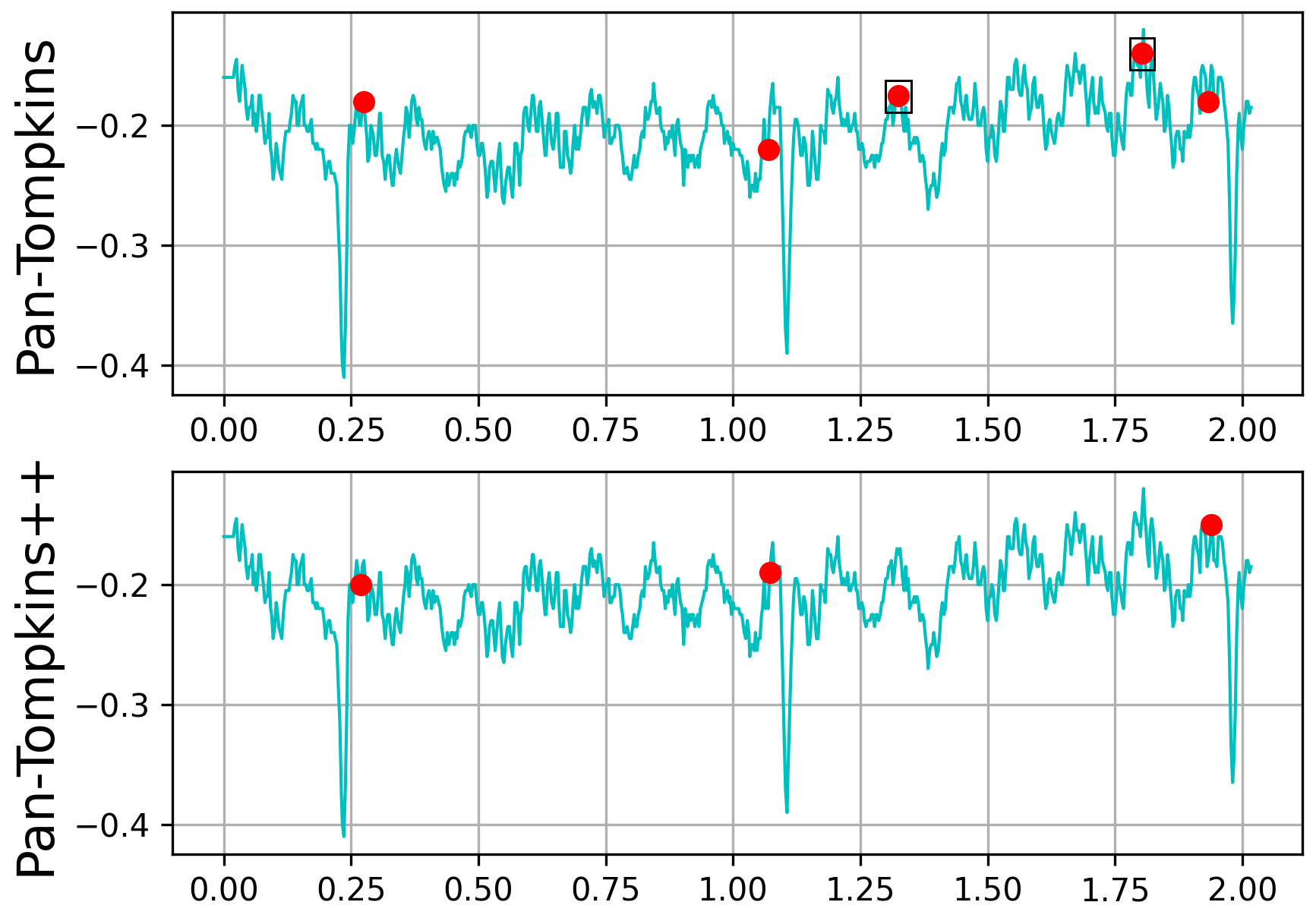}
    \caption{False R-peaks within the 360ms window of a beat (highlighted by black rectangles)  by Pan-Tompkins and proposed solution by Pan-Tompkins++.}
\end{center}
\label{fig3}
\end{figure}

\textbf{Limitation 5: No R-peak in a long period of time:}
\label{limitation5}
The Pan-Tompkins algorithm occasionally fails to detect R-peaks for a long period of time. Most of the time, it happens when there is a change in the signal pattern. Figure 4 shows the False Negatives given by the Pan-Tompkins algorithm in a defined interval. It ignores several actual R-peaks (marked with a black rectangle) as it does not take the nearby pattern into account. 

\textbf{Proposed solution for limitation 5:}
\label{solution5}
Our proposed algorithm minimizes the False Negatives by searching back in a predefined interval. If a QRS complex is not found during a defined interval (166 percent of the current RR-interval average or 1s window after the most recent detected QRS complex) a search-back operation is conducted to find if there is any missing QRS complex over that duration. $Threshold_3$ is calculated as:
\begin{align}
Threshold_3=0.5 \ Threshold_2 + 0.5 \ MEANSB
\end{align}
where, $MEANSB$ is a window with the preceding 3 QRS complexes and the following 3 peaks.

The search-back operation scans the window, starting after 360ms of the most recent R-peak up to the current peak, to find the probable missing R-peak. If the maximal point in that window is greater than $Threshold_3$, it is classified as an R-peak. When the QRS complex is found using the third threshold, the $SPK$ and $NPK$ are adjusted using $Rule-2$ (eqs. 12 and 13).
\begin{figure}[htbp]
\begin{center}
    \includegraphics[height=6cm,width=8cm]{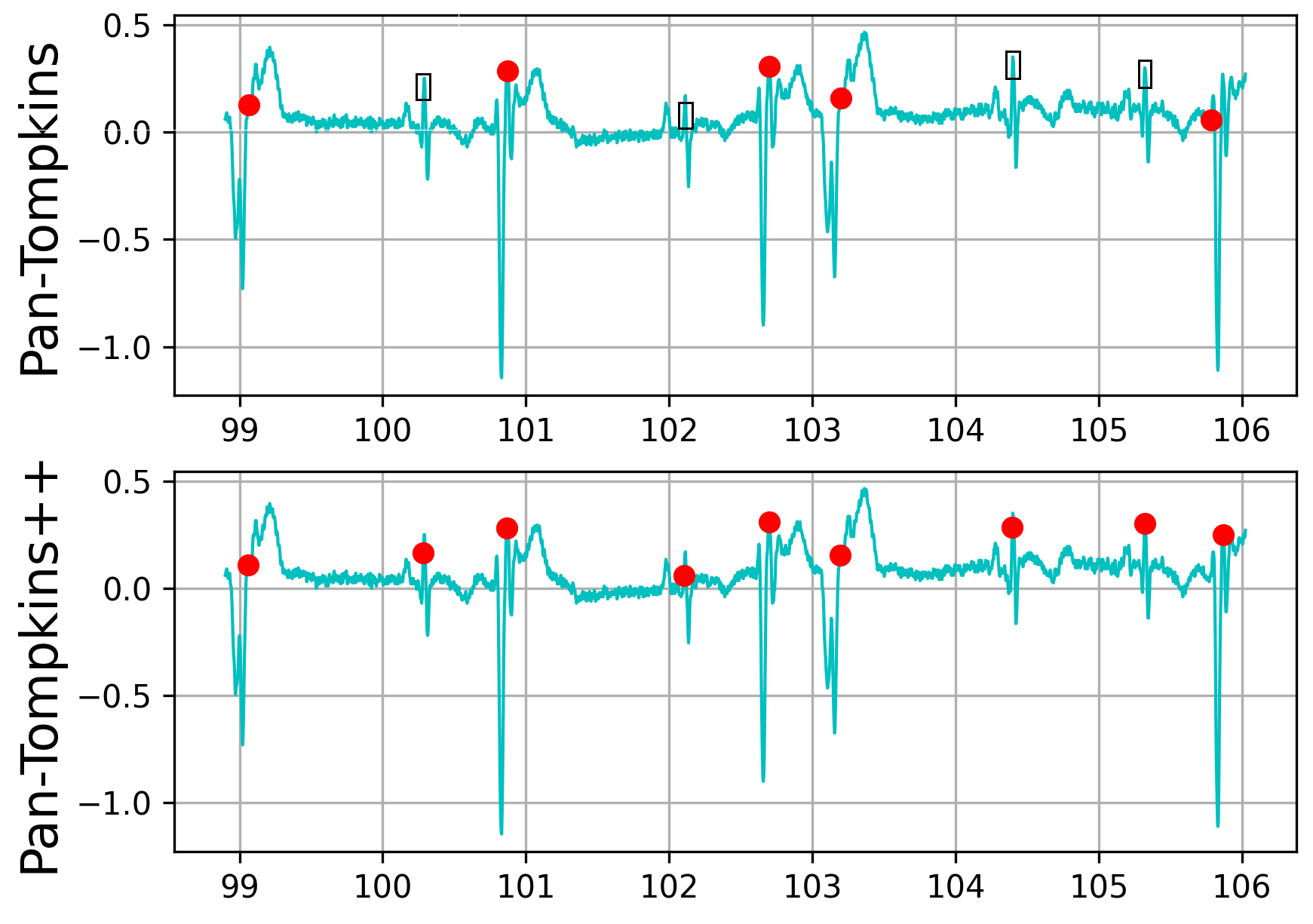}
    \caption{Missing R-peaks in the defined interval (highlighted by black rectangles) and effect of proposed solution.}
\end{center}
\label{fig4}
\end{figure}

\begin{figure}[htbp]
\begin{center}
    \includegraphics[height=6cm,width=8cm]{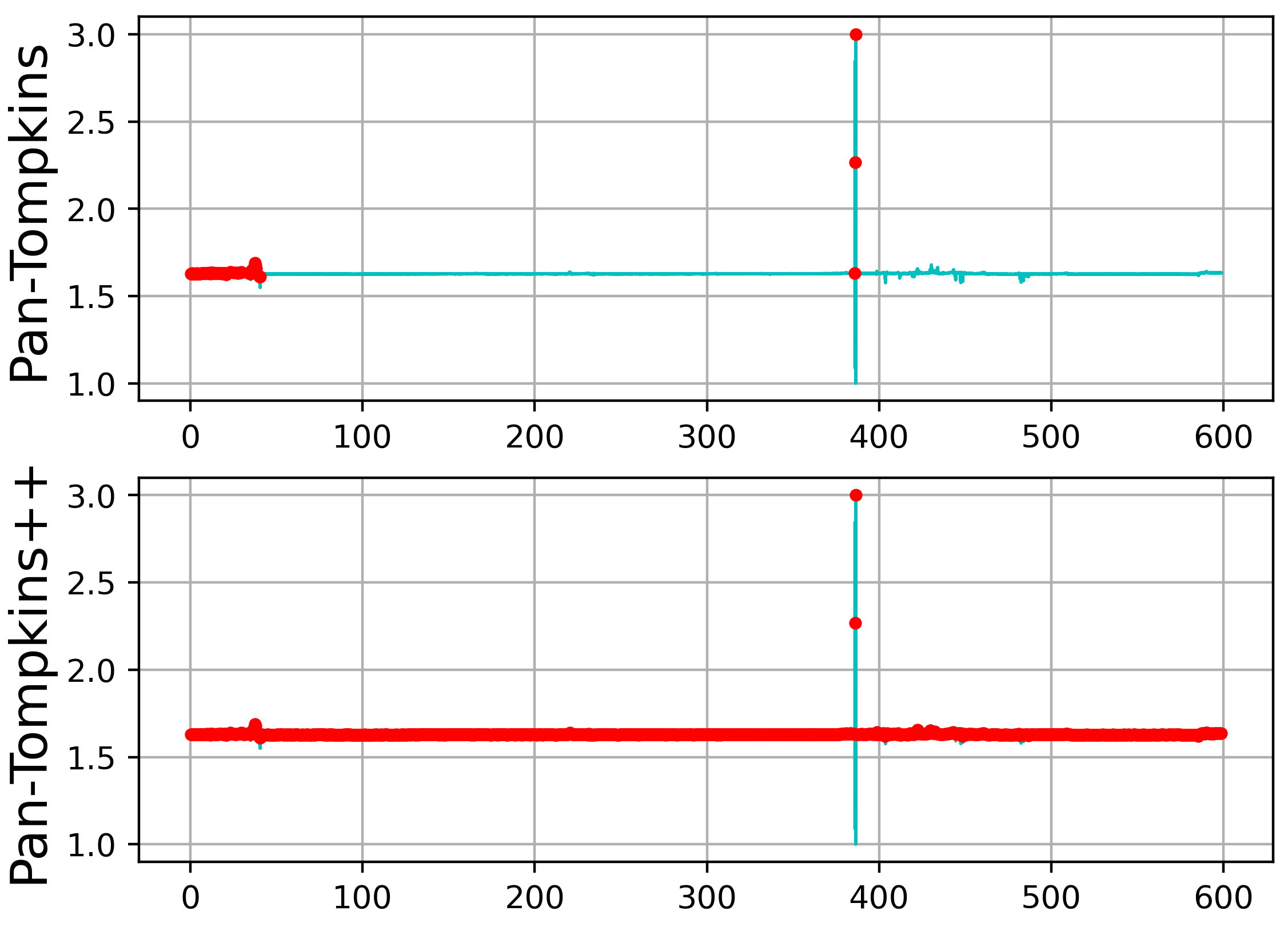}
    \caption{Missing R-peaks after exceptionally high signal components by Pan-Tompkins and proposed solution. This is a zoomed out version of the signal to show the effect on an entire 600s ECG signal. The entire ``red bar'' in Pan-Tompkins++ represents many R-peaks that were missed by Pan-Tompkins.}
\end{center}
\label{fig5}
\end{figure}

\textbf{Limitation 6: No R-peak after an exceptionally high R-peak:}
\label{limitation6}
The Pan-Tompkins algorithm struggles when it finds a signal component that is exceptionally high in comparison to the regular pattern. As the threshold adjustment is partially dependent on the current peak, the thresholds become quite high and the algorithm fails to detect QRS complexes afterward.  Figure 5 illustrates such a scenario. 

\textbf{Proposed solution for limitation 6:}
\label{solution6}
Our proposed algorithm provides a solution to this problem. If a QRS complex is not found during recent 1.4s, the probability of a missing QRS complex is quite high. When such an event occurs, the algorithm searches the window, starting after 360ms of the most recent R-peak up to the current peak, for the likely missing R-peak. An R-peak is determined as the maximum point in that window that is greater than 20 percent of the second threshold. When the QRS complex is found using this condition, the $SPK$ and $NPK$ are adjusted using $Rule-2$ (eqs. 12 and 13). Figure 5 shows that after encountering an exceptionally high signal component, Pan-Tompkins fails to detect R-peaks for a long period of time until it finds another exceptionally high R-peak. Pan-Tompkins could not detect any R-peaks in the duration of 41s–385s and in the duration of 387s–600s. But Pan-Tompkins++ successfully identifies R-peaks in that scenario.

The Pseudo-code of the Pan-Tompkins++ algorithm is presented in Table I. 

\begin{table}[htbp]
\begin{center}
\caption {Pseudo-code of the Pan-Tompkins++ algorithm}
\begin{tabular}{l*{8}{c}r}
\hline
\emph{\textbf{Preprocessing Phase}}\\
\  \emph{Apply a bandpass filter with a passband of 5–18 Hz to}\\
\ \ \ \emph{remove the noise in ECG} \\
\  \emph{Differentiate the signal}\\
\  \emph{Perform point by point squaring}\\
\  \emph{Smooth the signal using a 60ms wide flattop window}\\ 
\ \ \ \emph{function}\\
\  \emph{Apply moving window integration with a 150ms wide}\\
\ \ \ \emph{window to extract waveform feature information} \\
\emph{\textbf{Decision Phase}}\\
\ \emph{Select all the peaks that are at least 231ms apart from}\\ \ \ \ \emph{each other}\\
\ \emph{Initialize the thresholds}\\
\ \emph{\textbf{for }each Peak:}\\
\ \ \ \emph{\textbf{if} $Peak$$>$$Threshold_1$:}\\
\ \ \ \ \ \emph{Classify it as R-peak}\\
\ \ \ \ \ \emph{Adjust SPK and NPK using Rule-1}\\
\ \ \ \emph{\textbf{if} number of detected beats$>$8:}\\
\ \ \ \ \ \emph{Calculate the Mean RR-interval of the eight most}\\
\ \ \ \ \ \ \ \emph{recent beats}\\
\ \ \ \ \ \emph{Calculate current RR-interval}\\
\ \ \ \ \ \emph{\textbf{if} RR-interval$<$360ms or RR-interval$<$0.5 *}\\
\ \ \ \ \ \emph{Mean RR-interval:}\\
\ \ \ \ \ \ \ \emph{Classify it as T-wave or QRS complex based on}\\ 
\ \ \ \ \ \ \ \ \ \emph{slope}\\
\ \ \ \ \ \ \ \emph{Adjust SPK and NPK using Rule-1, if QRS complex}\\ 
\ \ \ \ \ \ \ \ \ \emph{found}\\
\ \ \ \ \ \emph{\textbf{else if} RR-interval$>$1s or RR-interval $>$1.66 * Mean }\\
\ \ \ \ \ \emph{RR-interval:}\\
\ \ \ \ \ \ \ \emph{$Threshold_3$=0.5 * $Threshold_2$ + 0.5 * MEANSB}\\ \ \ \ \ \ \ \ \emph{\textbf{if} Peak$>$$Threshold_3$:}\\
\ \ \ \ \ \ \ \ \ \emph{Classify it as R-peak}\\
\ \ \ \ \ \ \ \ \ \emph{Adjust SPK and NPK using Rule-2}\\
\ \ \ \ \ \emph{\textbf{else if} RR-interval$>$1.4s:}\\
\ \ \ \ \ \ \ \emph{\textbf{if} Peak$>$0.2 * $Threshold_2$}\\
\ \ \ \ \ \ \ \ \ \emph{Classify it as R-peak}\\
\ \ \ \ \ \ \ \ \ \emph{Adjust SPK and NPK using Rule-2}\\
\ \ \ \emph{$Threshold_1$ = NPK + 0.25 (SPK-NPK)}\\
\ \ \ \emph{$Threshold_2$ =0.4 * $Threshold_1$}\\
\hline
\end{tabular}
\label{tab1}
\end{center}
\end{table}

\section{Experimental Results}
\label{results}
The proposed 	Pan-Tompkins++ algorithm has been tested on four open-source ECG datasets. The performance of the proposed algorithm has been compared with the Pan-Tompkins algorithm as well as four other popular algorithms. The tolerance window (maximum offset from the annotation for a detected R-peak to be considered correct) is set at 100ms throughout the evaluation.

Although some recent machine learning and transform-based algorithms report results on some of these datasets, they use a reduced set of records \cite{gupta2021r}\cite{yuen2019inter} or a different tolerance window size\cite{chen2020automatic}\cite{zhao2018multilead}. The slow execution time and high computational requirement for training is likely the reason for using a reduced set of records. We only compare with signal processing-based algorithms, since they are still the most practical and widely utilized across the healthcare industry. 
\\
\subsubsection{Evaluation Metrics}
\label{metrics}
The performance of the proposed algorithm has been evaluated by the standard evaluation metrics, such as Positive Predictive Value (PPV), Sensitivity, F-score, and Execution Time. The equations for measuring these metrics are stated below.
$$PPV= \frac{\sum_{i=1}^{n}TP_i}{\sum_{i=1}^{n}TP_i+FP_i}$$ 
$$Sensitivity= \frac{\sum_{i=1}^{n}TP_i}{\sum_{i=1}^{n}TP_i+FN_i}$$
$$F-Score= \frac{2\times PPV \times Sensitivity}{PPV+Sensitivity}$$

$TP$, $FP$, $FN$, and $n$ are True Positive, False Positive, False Negative, and total records, respectively.  

\begin{table}[!htbp]
\caption {Performance comparison between the proposed Pan-Tompkins++ algorithm and selected R-peak detection algorithms on MIT-BIH Arrhythmia Dataset}
\begin{center}

\begin{tabular}
{p{0.19\linewidth} p{0.06\linewidth} p{0.06\linewidth} p{0.07\linewidth} p{0.1\linewidth} p{0.1\linewidth} p{0.1\linewidth}}
 \hline 
     \multicolumn{7}{c}{\textbf{MIT-BIH Arrhythmia Dataset (II)}} \\ 
     \hline
     
 & FP  (\%) & FN  (\%) & PPV  (\%) & Sensiti-vity(\%) & F-score(\%) & Execution Time(s)\\
\hline
Pan-Tompkins++ & 	\textbf{0.38}	& \textbf{0.51}	& \textbf{99.60}	& \textbf{99.47}	& \textbf{99.54}	& \textbf{47}\\
\hline
Pan-Tompkins\cite{pan1985real} &	0.83	& 0.81	& 99.15	& 99.17	& 99.16	& 72\\
\hline
Hamilton\cite{hamilton2002open} &	0.57	& 1.48	& 99.41	& 98.48	& 98.94	& 197\\
\hline
Modified Engzee\cite{engelse1979single}\cite{lourencco2012real} 	& 1.10	& 2.95	& 98.84	& 96.97	& 97.90	& 289\\
\hline
Wavelet Transform\cite{kalidas2017real}	& 1.15	& 1.73	& 98.81	& 98.22	& 98.51	& 321\\
\hline
Two Moving Averages\cite{elgendi2010frequency}	& 3.54	& 2.11	& 96.41	& 97.83	& 97.12	& 714\\
\hline
\end{tabular}

\begin{tabular}
{p{0.19\linewidth} p{0.06\linewidth} p{0.06\linewidth} p{0.07\linewidth} p{0.1\linewidth} p{0.1\linewidth} p{0.1\linewidth}}
 \hline 
     \multicolumn{7}{c}{\textbf{MIT-BIH Arrhythmia Dataset (V5)}} \\ 
     \hline
     
Pan-Tompkins++ & 	\textbf{4.16}	& \textbf{4.52}	& \textbf{95.71}	& \textbf{95.35}	& \textbf{95.53}	& \textbf{52}\\
\hline
Pan-Tompkins\cite{pan1985real} &	6.58 &	7.34 &	93.81 &	92.44 &	92.81 &	131\\
\hline
Hamilton\cite{hamilton2002open} &	5.76 &	8.33 &	93.91 &	91.43 &	92.65 &	114\\
\hline
Modified Engzee\cite{engelse1979single}\cite{lourencco2012real} 	& 4.31 & 5.60 &	95.50 &	94.24 &	94.87 &	289\\
\hline
Wavelet Transform\cite{kalidas2017real}	& 4.33 & 11.03 & 95.22 & 88.65 & 91.82 & 241\\
\hline
Two Moving Averages\cite{elgendi2010frequency}	& 10.43	& 7.89	& 89.54	& 91.88	& 90.69	& 583\\
\hline
\end{tabular}
\label{tab2}
\end{center}
\end{table}
The MIT-BIH Arrhythmia Dataset has 48 half-hour excerpts of two-channel (II, V5) ambulatory ECG recordings. The recordings are from 47 subjects taken by the BIH Arrhythmia Laboratory between 1975 and 1979. For both channels, our proposed algorithm outperforms all the other algorithms (Table II). The Pan-Tompkins++ yields the best outcomes for all performance matrices. In addition to improving Pan-Tompkins' R-peak detection accuracy, the proposed algorithm greatly reduces the execution time. A significant performance improvement is observed from the Pan-Tompkins algorithm for channel-V5. Several false beats are detected by the Pan-Tompkins algorithm within a very short period of time from the actual beats in ECG records 101, 102, 104, 105, 106, 203, 209, and 210. Pan-Tompkins++ avoids most of them because it distinguishes between the T-wave and QRS complex based on the slope information when it finds any peak within 360ms or within 50\% of the mean RR-interval from the true R-peak. 

\begin{table}[htbp]
\caption {Performance comparison between the proposed Pan-Tompkins++ algorithm and selected R-peak detection algorithms on St.-Petersburg Institute of Cardiological Technics 12-lead Arrhythmia Dataset}
\begin{center}

\begin{tabular}
{p{0.19\linewidth} p{0.06\linewidth} p{0.06\linewidth} p{0.07\linewidth} p{0.1\linewidth} p{0.1\linewidth} p{0.1\linewidth}}
 \hline 
     \multicolumn{7}{c}{\textbf{St.-Petersburg Institute of Cardiological Technics 12-lead}} \\ 
     
     \multicolumn{7}{c}{\textbf{Arrhythmia Dataset (I)}} \\ 
     \hline
     
 & FP  (\%) & FN  (\%) & PPV  (\%) & Sensiti-vity(\%) & F-score(\%) & Execution Time(s)\\
\hline
Pan-Tompkins++	& \textbf{3.91} & \textbf{4.11} &	\textbf{96.08} &	\textbf{95.89} &	\textbf{95.98} &	\textbf{94}\\
\hline
Pan-Tompkins\cite{pan1985real} &	7.11 & 6.33 & 92.95 &	93.67 &	93.31 &	189\\
\hline
Hamilton\cite{hamilton2002open} &	5.76 &	8.77 &	94.06 &	91.23 &	92.63 &	198\\
\hline
Modified Engzee\cite{engelse1979single}\cite{lourencco2012real} &	4.66 &	8.66 &	95.15 &	91.34 &	93.20 &	324\\
\hline
Wavelet Transform\cite{kalidas2017real} &	8.28 &	16.0 &	91.03 &	84.00 &	87.37 &	272\\
\hline
Two Moving Averages\cite{elgendi2010frequency}	& 7.29 &	6.51 &	92.76 &	93.49 &	93.13 &	627\\
\hline
\end{tabular}\\

\begin{tabular}
{p{0.19\linewidth} p{0.06\linewidth} p{0.06\linewidth} p{0.07\linewidth} p{0.1\linewidth} p{0.1\linewidth} p{0.1\linewidth}}

 \hline 
     \multicolumn{7}{c}{\textbf{St.-Petersburg Institute of Cardiological Technics 12-lead}} \\ 

     \multicolumn{7}{c}{\textbf{Arrhythmia Dataset (AVR)}} \\ 
     \hline
Pan-Tompkins++ &	\textbf{0.80} &	0.89 &	\textbf{99.20} &	99.11 &	\textbf{99.15} &	\textbf{75}\\
\hline
Pan-Tompkins\cite{pan1985real} &	1.68 &	\textbf{0.62} &	98.34 &	\textbf{99.38} &	98.86 &	99\\
\hline
Hamilton\cite{hamilton2002open} &	1.48 &	2.50 &	98.50 &	97.50 &	98.00 &	171\\
\hline
Modified Engzee\cite{engelse1979single}\cite{lourencco2012real} &	1.56 &	12.19 &	98.25 &	87.81 &	92.74 &	329\\
\hline
Wavelet Transform\cite{kalidas2017real} &	2.11 &	6.45 &	97.62 &	93.54 &	95.62 &	278\\
\hline
Two Moving Averages\cite{elgendi2010frequency}	& 3.39 & 2.37 &	96.65 &	97.63 &	97.13 &	555\\
\hline
\end{tabular}\\

\begin{tabular}
{p{0.19\linewidth} p{0.06\linewidth} p{0.06\linewidth} p{0.07\linewidth} p{0.1\linewidth} p{0.1\linewidth} p{0.1\linewidth}}

  \hline 
     \multicolumn{7}{c}{\textbf{St.-Petersburg Institute of Cardiological Technics 12-lead}} \\ 

     \multicolumn{7}{c}{\textbf{Arrhythmia Dataset (V2)}} \\ 
     \hline

Pan-Tompkins++ &	\textbf{0.12} &	0.33 &	\textbf{99.88} &	99.67 &	\textbf{99.77} &	\textbf{76}\\
\hline
Pan-Tompkins\cite{pan1985real} &	0.23 &	\textbf{0.32} &	99.76 &	\textbf{99.68} &	99.72 &	96\\
\hline
Hamilton\cite{hamilton2002open} &	0.38 &	1.07 &	99.62 &	98.93 &	99.27 &	214\\
\hline
Modified Engzee\cite{engelse1979single}\cite{lourencco2012real} &	0.27 &	2.25 &	99.73 &	97.75 &	98.73 &	331\\
\hline
Wavelet Transform\cite{kalidas2017real} &	0.46 &	1.06 &	99.54 &	98.93 &	99.24 &	273\\
\hline
Two Moving Averages\cite{elgendi2010frequency} &	1.40 &	0.94 &	98.60 &	99.06 &	98.83 &	543\\
\hline
\end{tabular}
\label{tab3}
\end{center}
\end{table}

The St.-Petersburg Institute of Cardiological Technics 12-lead Arrhythmia Dataset has 75 annotated recordings that were taken from 32 Holter records. Each record is 30 minutes long and has 12 standard leads. Among the 12 leads, three (I, AVR, V2) have been chosen for investigation. Table III shows the results given by the algorithms for the records from three leads. It is clear that the Pan-Tompkins++ has a higher performance than the classical methods for all three lead recordings. For lead-I, the PPV, sensitivity, and F-score increase notably with a comparison of the Pan-Tompkins algorithm while the execution time decreases drastically from 189s to 94s. Overall, a performance improvement is noticed for lead-V2 except for the sensitivity, which drops by 0.01\%.

\begin{table}[htbp]
\caption {Performance comparison between the proposed Pan-Tompkins++ algorithm and selected R-peak detection algorithms on European ST-T Dataset}
\begin{center}
\begin{tabular}
{p{0.19\linewidth} p{0.06\linewidth} p{0.06\linewidth} p{0.07\linewidth} p{0.1\linewidth} p{0.1\linewidth} p{0.1\linewidth}}

  \hline 
     \multicolumn{7}{c}{\textbf{European ST-T Dataset (V4)}} \\ 
     \hline
     
 & FP  (\%) & FN  (\%) & PPV  (\%) & Sensiti-vity(\%) & F-score(\%) & Execution Time(s)\\
\hline
Pan-Tompkins++ & \textbf{2.44} &	\textbf{2.58} &	\textbf{97.56} &	\textbf{97.42} &	\textbf{97.49} &	\textbf{129}\\
\hline
Pan-Tompkins\cite{pan1985real} &	6.07 &	4.14 &	94.04 &	95.86 &	94.94 &	145\\
\hline
Hamilton\cite{hamilton2002open} &	4.56 &	4.99 &	95.42 &	95.01 &	95.21 &	341\\
\hline
Modified Engzee\cite{engelse1979single}\cite{lourencco2012real} &	3.39 &	4.39 &	96.57 &	95.61 &	96.09 &	557\\
\hline
Wavelet Transform\cite{kalidas2017real} &	3.24 &	3.51 &	96.75 &	96.49 &	96.62 &	482\\
\hline
Two Moving Averages\cite{elgendi2010frequency}	& 5.84 & 4.05 &	94.26 &	95.95 &	95.10 &	942\\
\hline
\end{tabular}\\
\begin{tabular}
{p{0.19\linewidth} p{0.06\linewidth} p{0.06\linewidth} p{0.07\linewidth} p{0.1\linewidth} p{0.1\linewidth} p{0.1\linewidth}}

  \hline 
     \multicolumn{7}{c}{\textbf{European ST-T Dataset (III)}} \\ 
     \hline
    
Pan-Tompkins++ & \textbf{1.59} &	\textbf{1.02} &	\textbf{98.42} &	\textbf{98.98} &	\textbf{98.70} &	\textbf{138}\\
\hline
Pan-Tompkins\cite{pan1985real} &	6.89 &	6.54 &	93.13 &	93.46 &	93.30 &	141\\
\hline
Hamilton\cite{hamilton2002open} &	3.60 &	3.95 &	96.38 &	96.05 &	96.21 &	319\\
\hline
Modified Engzee\cite{engelse1979single}\cite{lourencco2012real} &	4.13 &	4.71 &	95.84 &	95.29 &	95.57 &	565\\
\hline
Wavelet Transform\cite{kalidas2017real} &	4.14 &	5.70 &	95.79 &	94.30 &	95.04 &	471\\
\hline
Two Moving Averages\cite{elgendi2010frequency} & 5.12 & 3.71 &	94.95 &	96.29 &	95.62 &	893\\
\hline
\end{tabular}
\label{tab4}
\end{center}
\end{table}

The European ST-T Dataset consists of 90 two hour annotated records of two-channel (III, V4) from 79 subjects. The first 40 minutes of each record have been taken for evaluation as it would take a long time to evaluate the entire record. From the Pan-Tompkins algorithm, a major improvement is observed with the proposed algorithm for records from both of the channels (Table IV). For lead-III, the PPV, sensitivity, and F-score increase by 5.29, 5.52, and 5.4 percent, respectively. In records 116, 119, 205, 206, 405, 602, 605, 611, 613, and 817, the Pan-Tompkins algorithm fails to detect R-peaks when the signal changes its usual pattern. Pan-Tompkins++ detects those R-peaks as it considers the preceding and following signal components from the current position. 

The Computing in Cardiology Challenge 2014 Dataset consists of 100 ten-minute annotated recordings from lead II. Here, all the algorithms perform poorly when compared to the other datasets. However, Pan-Tompkins++ outperforms all the other algorithms by a significant margin (Table V). There are some R-peaks with exceptionally high magnitude found in records 1028, 1522, 1683, 1715, 1804, 2201, 2203, 2370, and 2664. The Pan-Tompkins algorithm could not detect true R-peaks for a long period of time after those high R-peaks as the thresholds became high. Pan-Tompkins++ successfully detects those true R-peaks because it compares peaks with a reduced threshold when there is no R-peak detected in recent 1.4s. The improvement over the Pan-Tompkins algorithm is notable for PPV and execution time.

Figure 6 illustrates a situation where Pan-Tompkins++ (PT++) effectively detects the true R-peaks while all the other algorithms fail. Hamilton (Ham), Modified Engzee (ME), Wavelet Transform (WT), and Two Moving Averages (TMA) algorithms detect false R-peaks (blue x) as they find sharp transitions apart from the true QRS complex. Pan-Tompkins (PT) identifies both false (blue x) and true (red circle) R-peaks. Pan-Tompkins++ (PT++) successfully detects only the real R-peaks (red circle) as it compares the morphology of the current candidate QRS complex with that of the most recent detected beat over a 70ms window, and there is a 231ms refractory period once a valid QRS complex is found.

 Pan-Tompkins++ outperforms all the selected classical algorithms for each of the four datasets. There is a significant reduction in execution time, which makes the algorithm more suitable for real-time heart monitoring applications. One of the reasons for the reduction of execution time is that there is a 231ms refractory period following the identification of a valid QRS complex before the next one can be found, as we allow a maximum heart rate of 260 bpm. It is 200ms in the Pan-Tompkins algorithm. Moreover, in the Pan-Tompkins algorithm, the thresholds are adjusted whenever the current RR-interval is less than 92\% or greater than 116\% of the mean RR-interval. This adjustment has been removed as there was no improvement found with it.

It is important to note that although the performance of these algorithms seem high, there is still room for improvement, even for Pan-Tompkins++. Even a small number of false or missed detections can result in misdiagnosis or cardiac events with catastrophic consequences.

\begin{table}[!htbp]
\caption {Performance comparison between the proposed Pan-Tompkins++ algorithm and selected R-peak detection algorithms on Computing in Cardiology Challenge 2014 Dataset}
\begin{center}
\begin{tabular}
{p{0.19\linewidth} p{0.06\linewidth} p{0.06\linewidth} p{0.07\linewidth} p{0.1\linewidth} p{0.1\linewidth} p{0.1\linewidth}}

  \hline 
     \multicolumn{7}{c}{\textbf{Computing in Cardiology Challenge 2014 Dataset}} \\ 
     \hline
     
 & FP  (\%) & FN  (\%) & PPV  (\%) & Sensiti-vity(\%) & F-score(\%) & Execution Time(s)\\
\hline
Pan-Tompkins++ & \textbf{18.97} & \textbf{17.64} & \textbf{81.28} & \textbf{82.36} & \textbf{81.81} & \textbf{49}\\
\hline
Pan-Tompkins\cite{pan1985real} &	25.01 &	19.74 &	76.24 &	80.26 &	78.20 &	82\\
\hline
Hamilton\cite{hamilton2002open} &	25.37 &	27.56 &	74.07 &	72.44 &	73.24 &	131\\
\hline
Modified Engzee\cite{engelse1979single}\cite{lourencco2012real} &	25.39 &	23.11 &	75.18 &	76.89 &	76.02 &	187\\
\hline
Wavelet Transform\cite{kalidas2017real} &	23.89 &	34.88 &	73.16 &	65.12 &	68.91 &	134\\
\hline
Two Moving Averages\cite{elgendi2010frequency} & 25.09 & 20.27 & 76.07 & 79.73 & 77.86 & 265\\
\hline
\end{tabular}
\label{tab5}
\end{center}
\end{table}

\begin{figure}[htbp]
\begin{center}
    \includegraphics[height=6cm,width=8cm]{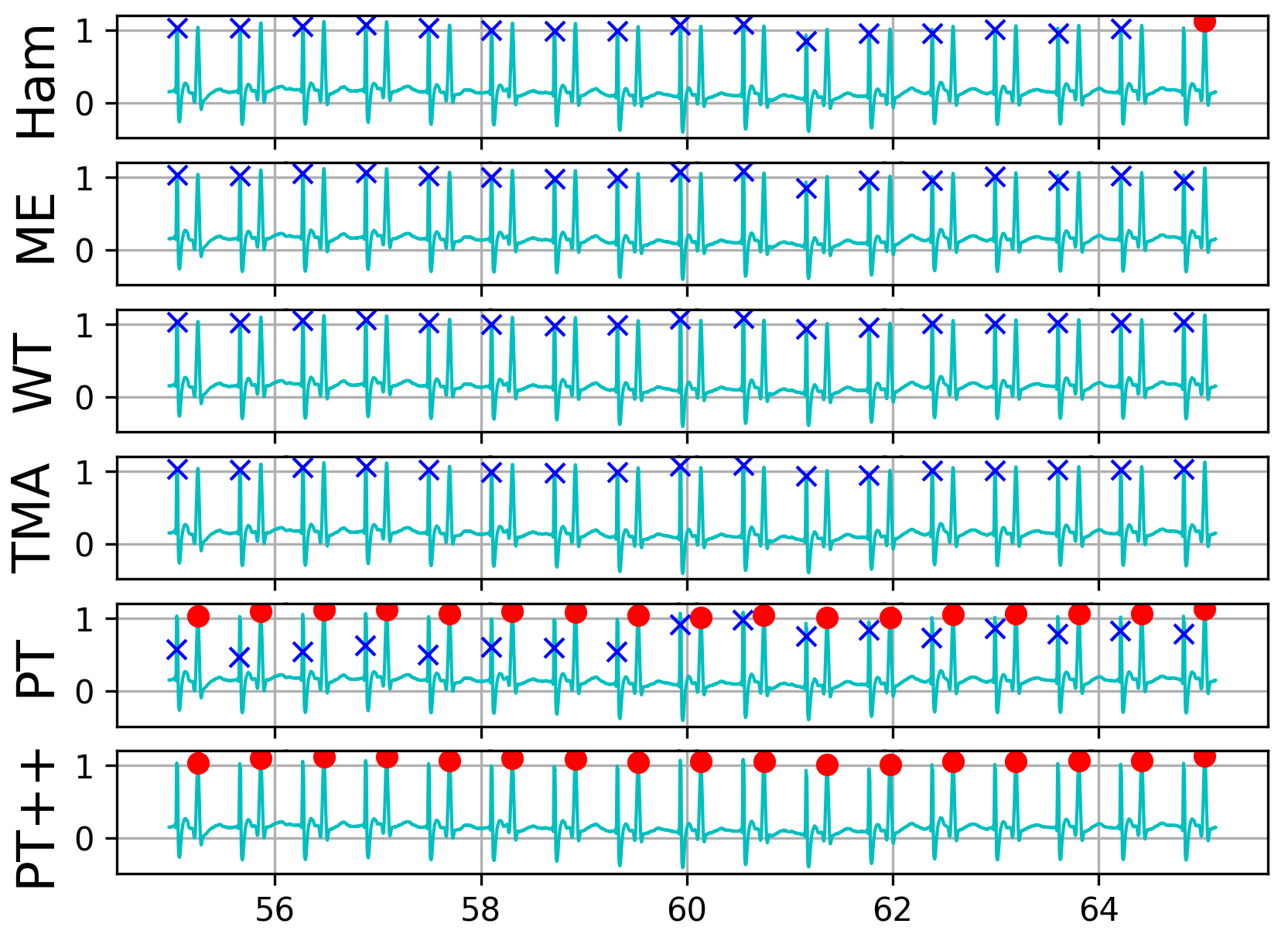}
    \caption{Successful R-peak detection by Pan-Tompkins++ where others fail. blue x- false positive, red circle - true positive.}
\end{center}
\label{fig6}
\end{figure}
\section{Conclusion}
\label{conclusion}
In this paper, we presented Pan-Tompkins++, inspired by the Pan-Tompkins algorithm, for better R-peak detection in ECG. The preprocessing and thresholding approaches were chosen to enhance true R-peak detection. This would do away with the requirement for a separate filtering function altogether. The proposed algorithm greatly improves the performance of the Pan-Tompkins algorithm. It effectively reduces the False Positives and False Negatives given by the Pan-Tompkins algorithm and thus lessens the algorithm’s drawbacks. Additionally, it significantly cuts down on execution time, making it better suited for real-time cardiac monitoring applications. Obtained results using ECGs from four open source datasets as evaluation sets demonstrate that more noteworthy performances are found in comparison with selected well-known algorithms. In our future work, we intend to further validate the suggested algorithm with real-life ECG data.
\section*{Acknowledgment}
The authors acknowledge funding from Myant Inc. through the Mitacs Canada Accelerate program. 
\bibliography{ref} 
\bibliographystyle{IEEEtran}
\end{document}